\documentclass[pra,twocolumn,epsfig,rotate,showpacs]{revtex4}
\usepackage{epsfig}
\usepackage{graphicx}
\usepackage{amsmath}

\newcommand{\beq}{\begin{eqnarray}}
\newcommand{\eeq}{\end{eqnarray}}
\begin{document}
\title{Adiabatic Quantum Transport in a Spin Chain with a
Moving Potential}
\author{Vinitha Balachandran
and Jiangbin Gong} \email{phygj@nus.edu.sg}\affiliation{Department
of Physics and Centre for Computational Science and Engineering, \\
National University of Singapore, 117542, Republic of Singapore}

\begin{abstract}
Many schemes to realize quantum state transfer in spin chains are
not robust to random fluctuations in the spin-spin coupling
strength. In efforts to achieve robust quantum state transfer, an
adiabatic quantum population transfer scheme is proposed in this
study. The proposed scheme makes use of a slowly moving external
parabolic potential and is qualitatively explained in terms of the
adiabatic following of a quantum state with a moving separatrix
structure in the classical phase space of a pendulum analogy.
Detailed aspects of our adiabatic population transfer scheme,
including its robustness, is studied computationally. Applications
of our adiabatic scheme in quantum information transfer are also
discussed, with emphasis placed on the usage of a dual spin chain to
encode quantum phases. The results should also be useful for the
control of electron tunneling in an array of quantum dots.
\end{abstract}
\date{\today}
\maketitle

\section{Introduction}
Efficiently transferring quantum information is an important
challenge for the practical realization of quantum computers.
Optical fibers, where quantum information is transmitted by mobile
carriers like photons, are most desirable for long-distance
communication. However, the need to interface with solid state
quantum computer components considerably restricts the experimental
feasibility of using optical systems as quantum channels. One
promising alternative is to use condensed matter systems themselves
as the {\it quantum wire} for information transfer. This idea,
initially advocated and studied by Bose \cite{bose1} in the context
of quantum spin chains, has now attracted wide interests. In
particular, for a quantum spin chain used as the quantum wire, the
natural evolution of permanently coupled spins is exploited to
accomplish the quantum information transfer. Because the
Hamiltonians equivalent to that of a spin chain may be realized in
different physical systems (e.g., arrays of Josephson junctions
\cite{plenio}, cold atoms in optical lattices \cite{duan03}, arrays
of quantum dots, etc.), interests in spin chains as a promising
candidate of quantum wires continue to grow.  Experimental studies
on the dynamics of quantum information transfer and on the quantum
control in spin chains using nuclear magnetic resonance techniques
have also been reported \cite{NMR}.

Despite many fruitful studies of spin chains from a quantum
information perspective, many theoretical problems are still open.
To be more specific let us consider first Bose's original proposal
\cite{bose1}. Therein the fidelity of quantum information transfer
is gradually degraded by the dispersion effects associated with the
quantum
propagation. 
Furthermore, particular external magnetic field should be designed
to ensure a correct quantum phase at the receiver's site. These
issues and others motivated a series of sophisticated protocols to
achieve better quantum information transfer. One noteworthy approach
was to pre-engineer the nearest-neighbor couplings of a spin chain
or even a spin network
\cite{christandl1,twamley,network2,stolze,superconducting}. A second
approach exploits the mirror symmetry of a spin chain
\cite{stolze,parabolic}. Another approach suggests to use Gaussian
wavepackets with slow dispersion \cite{osborne1} to encode the
quantum information to be transferred along an unmodulated spin
chain. Unfortunately, these pioneering approaches rely upon specific
analytical forms of the involved Hamiltonian and hence are not
robust to imperfection or physical fluctuations in the spin-spin
coupling strength. One encounters the same situation when applying
other more subtle techniques \cite{wojcik1,lyakhov1,hase}.

To have the desired robustness that may be necessary for any type of
quantum information transfer, two quantum control schemes based on
adiabatically varying the coupling strength in a spin chain have
also been suggested \cite{adiabatic1,adiabatic2}. Note however,
these schemes require individual addressing of the nearest-neighbor
coupling and hence present new experimental challenges.  Another
novel and quite robust quantum transfer protocol in spin chains is
the dual spin chain scheme
\cite{burgarth2,burgarth1,burgarth3,burgarthnj}. Therein the quantum
information to be transferred is encoded in two parallel spin
sub-chains (initially assumed to be identical, but even this
condition may be lifted under certain conditions). Thanks to the
encoding with two spin sub-chains, the quantum transfer can be very
robust to static disorder.  Nevertheless, even this promising scheme
is not perfect, because (i) it may need too many quantum
measurements (or too many steps of ``trial and error"), (ii) it may
not operate well in the presence of time-dependent disorder, and
(iii) the effects of nonideal measurements are still under
investigation \cite{burgarth3}.

Hence, it remains an open question as to which quantum transfer
scheme will ultimately be adopted experimentally, with high fidelity
and low cost. It is our belief that in the end a combination of
several techniques may be able to offer the most powerful protocol
for quantum information transfer in solid state systems.


In this paper, we introduce an adiabatic transport scheme assisted
by a slowly moving external field applied to a spin chain. Thanks to
a pendulum analogy, the central idea can be understood in a very
simple manner. The essence is that when an external field is moving
slowly, the spin excitation may adiabatically follow the field under
certain conditions. During this process the quantum population of
spin excitation is transferred from one end of the spin chain to the
other end. As we show below in detail, this adiabatic scheme offers
a number of advantages: (i) it is highly robust, (ii) it can be
operated rather fast (in the absence of disorder, it may be as fast
as one tenth of the natural propagation speed of the spin chain),
(iii) the required field strength can be decreased by using
wavepackets as initial states, (iv) the transfer can be easily
stopped and relaunched, (v) the time of arrival of the quantum
population transfer to the last spin with high probability can be
easily predicted, and (vi) it can offer a promising means of quantum
information transfer when combined with the above-mentioned dual
spin chain scheme.

The outline of the paper is as follows. In Sec. \ref{pendulum}, we
explain the motivation and mechanism of our adiabatic population
transfer scheme by mapping the spin chain Hamiltonian to that of a
pendulum. Detailed computational results are presented in Sec. III.
The robustness of our scheme is studied in Sec. IV. In Sec. V,
 we discuss how our adiabatic population
transfer scheme, which does not yet take care of the quantum phases
(also essential for quantum information transfer), can be combined
with the dual spin chain scheme to offer a potentially powerful
approach for quantum information transfer.
We conclude this work in Sec. VI.

\section{Adiabatic Quantum Transport in Spin Chains: A Pendulum Perspective}
\label{pendulum}
 Consider a one-dimensional Heisenberg chain of $N+1$
spins subject to an external parabolic magnetic field. The
associated Hamiltonian is given by
\begin{eqnarray}
 H_{s}&=&-\frac{J}{2}\sum_{n=0}^{N-1}{\bf \sigma}_{n}\cdot {\bf \sigma}_{n+1}
 +
 \sum_{n=0}^{N}\frac{C}{2}(n-n_{0})^{2}\sigma_{n}^{z},
 \label{KReq}
 \end{eqnarray}
where ${\bf \sigma}\equiv (\sigma^{x},\sigma^{y},\sigma^{z})$ are
the Pauli
 matrices, $J$ the coupling strength between nearest neighbor
 spins, and $C$ is proportional to the magnetic dipole of the spins and
the amplitude of the parabolic field whose minimum is at site
$n_{0}$. Note that $n_{0}$ will be time-dependent in our control
scheme. Below we assume all the system parameters have been
appropriately scaled and take dimensionless values, with $J=1$,
$\hbar=1$ throughout. As such, the energy scale (e.g., the parameter
$C$) should be understood with respect to $J$, and the time scale
should be understood with respect to $\hbar/J$. Because the spin
chain Hamiltonian $H_{s}$ commutes with the total
 polarization $S_{z}\equiv \sum_{n=1}^{N}\sigma_{n}^{z}$, the dynamics of the spin
 chain
 preserves the total polarization. Here we
 restrict our analysis to the subspace of $S_{z}
 =1-N$, where in total only one spin is flipped.
 In this subspace the total state of the chain can be
written as
\begin{eqnarray}
  |\Psi(t)\rangle &=& \sum_{m=0}^{N}c_{m}(t)|\mathbf{m}\rangle,
\end{eqnarray}
where $|\mathbf{m}\rangle$ represents one basis state with a spin up
at the $m$th site and all other spins down. The complex coefficients
$c_{m}(t)$ are the probability amplitude.  Below we also shift the zero
of the energy scale such that if one spin is down, its interaction
with the external magnetic field contributes zero to the total
energy.

To understand the essence of the spin chain dynamics from a
semiclassical perspective, we now consider the large $N$ limit of
the spin chain. Denote $k$ as the quasi-momentum of a plane spin
wave and the $|k\rangle$ the associated eigenstate of the
quasi-momentum. Then the Hamiltonian in Eq. (\ref{KReq}) can be
rewritten as
\begin{eqnarray}
H_{s}= -J \int_{0}^{2\pi} \cos(k) |k\rangle\langle k| \nonumber \\
+\sum_{n} |{\bf n}\rangle\langle {\bf n}| \frac{C}{2}(n-n_{0})^{2}.
\label{H2}
\end{eqnarray}
This form can be further simplified in an operator form, i.e.,
\begin{eqnarray}
H_{s}= -J \cos(\hat{k}) +\frac{C}{2}(\hat{n}-n_{0})^{2}\label{H3},
\end{eqnarray}
where $\hat{k}|k\rangle =k |k\rangle$; $\hat{n}|{\bf n}\rangle=n
|\bf{n}\rangle$; $[\cos(\hat{k}), \hat{n}]=-i\sin(\hat{k})$, and
$[\sin(\hat{k}), \hat{n}]=i\cos(\hat{k})$.  The Hamiltonian in Eq.
(\ref{H3}) can now be easily recognized to be the Hamiltonian of a
quantum pendulum $H_{p}$ \cite{Boness}, with an effective Planck
constant $\sqrt{C}$. Specifically, with the mapping
$\hat{k}\rightarrow \hat{x}$, $\sqrt{C}\hat{n}\rightarrow \hat{p}$,
the spin chain Hamiltonian $H_{s}$ is mapped to
\begin{eqnarray}
H_{p}= -J \cos(\hat{x}) +\frac{1}{2}(\hat{p}-p_{0})^{2}\label{H4},
\end{eqnarray}
where $p_{0}=\sqrt{C}n_{0}$ and $[\cos(\hat{x}),
\hat{p}]=-i\sqrt{C}\sin(\hat{x})$.  The semiclassical Hamiltonian
for this quantum pendulum, i.e., $H_{p}^{c}=-J \cos(x)
+\frac{1}{2}(p-p_{0})^{2}$ is obtained by replacing $\hat{x}$ and
$\hat{p}$ with $c$-variables $x$ and $p$.

   \begin{figure}
  \begin{center}
  \epsfig{file=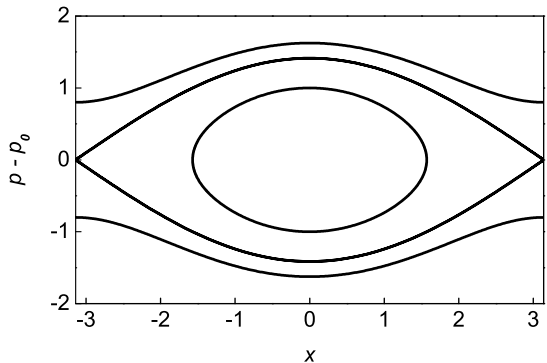,width=8.6cm}
  \end{center}
  \caption{Phase space portrait for the classical pendulum Hamiltonian associated with Eq. (\ref{H4}),
   with $C=2$ and $J=1$. The closed curve enclosing the island and intersecting with both
   $p-p_{0}=0$ and $x=\pm\pi$ is the separatrix.  Variables $P-P_{0}$ and
   $x$ plotted here take dimensionless values.
    Note that if the separatrix is moving up
   slowly by increasing $p_{0}$ gradually,
   an initial state enclosed by the separatrix is expected to adiabatically follow the movement of the separatrix.
  This suggests that a slowly moving parabolic potential applied to a spin chain can be used to
  adiabatically transfer spin excitation along a spin chain.}
  \label{separatrix}
\end{figure}

Many quantum transport features of the spin chain can now be
understood in terms of the semiclassical dynamics of the pendulum
analogy thus obtained. In particular, the quantum transport in the
momentum space of the pendulum is now in parallel with the transfer
of spin excitation from one site to another. Hence, the issue of
robust quantum transport of spin excitation along the spin chain now
reduces to the design of a control scenario that enables robust
transport of the pendulum state in its momentum space.

An adiabatic scheme for robust quantum population transfer along a
spin chain can now be proposed. The key observation is the existence
of a motional separatrix in the classical phase space of the
pendulum. This separatrix is located at $ -J \cos(x)
+(1/2)(p-p_{0})^{2}=J$ [see Fig. \ref{separatrix}]. If we now slowly
move up the separatrix along the momentum space by increasing $p_0$,
then a quantum state initially trapped inside the separatrix cannot
penetrate this separatrix and is expected to adiabatically follow
the moving separatrix, giving rise to adiabatic transport in the
momentum space. Translating this pendulum language back to the spin
chain case, one anticipates that a slowly moving parabolic magnetic
field (with slowly increasing $n_{0}$) should result in a robust
scenario for transferring quantum population along the spin chain.
During this process the dispersion of the spin wave should also be
bounded by the separatrix structure, i.e., a moving but
non-spreading wavepacket \cite{report} of the spin wave can be
expected.  The main remaining task of this paper is devoted to
detailed aspects of this adiabatic control scheme.

It is also interesting to note that the dynamics of the spin chain
can be mapped to that of a tight-binding model. To see this consider
first the associated Schr\"{o}dinger equation,
\begin{eqnarray}
    i\frac{dc_{0}}{dt} &=& \frac{J}{2}c_{1} + \frac{C}{2}n_{0}^{2}c_{0}, \nonumber  \\
    i\frac{dc_{n}}{dt}  &=& \frac{J}{2}(c_{n-1}+c_{n+1}) + \frac{C}{2}(n-n_{0})^{2}c_{n},\ 0<n<N, \nonumber \\
    i\frac{dc_{N}}{dt}  &=& \frac{J}{2}c_{N-1} +
    \frac{C}{2}(N-n_{0})^{2}c_{N}.
    \label{seq}
  \end{eqnarray}
Consider next a tight-binding Hamiltonian $H_{t}$ describing, for
example, an array of $(N+1)$ identical quantum dots subject to an
external parabolic field, with one electron tunneling between the
quantum dots. Then $H_{t}$ assumes the following form,
\begin{eqnarray}
  H_{t} &=&-\frac{J}{2}\sum_{n=0}^{N-1}(a_{n}^{\dagger}a_{n+1}
   +a_{n}a_{n+1}^{\dagger}) \nonumber \\
   && +\sum_{n=0}^{N}\frac{C}{2}(n-n_{0})^{2}a_{n}^{\dagger}a_{n},
\end{eqnarray}
where $J$ represents the constant tunneling rate between the
nearest-neighbor quantum dots, and $a_{n}^{\dagger}$ and $a_{n}$
represent the creation and annihilation operators.
Because the total number of electrons is already assumed to be one,
the system wavefunction can also be written as
  $|\Psi(t)\rangle = \sum_{m=0}^{N}c_{m}(t)|\mathbf{m}\rangle$,
where $|\mathbf{m}\rangle$  denotes the state with an electron in
the $m$th quantum dot and  $c_{m}(t)$ denotes the associated quantum
amplitude. In this representation, one immediately finds that the
evolution of this tight-binding system takes the same form as Eq.
(\ref{seq}). Thus, the above pendulum analogy is also applicable to
a tight-binding system and is hence very useful for consideration of
adiabatic quantum transport in quantum dot arrays \cite{adiabatic3}.
Note also that one may start from Eq. (\ref{seq}) to have an
alternative derivation of the pendulum analogy \cite{kol1}.

To end this section, we stress that the proposed control scheme is
based upon a semiclassical perspective afforded by the pendulum
analogy. What is not addressed is the issue of transferring the
quantum phase along the spin chain.  As such, although the pendulum
analogy helps design our scheme for the robust transport of quantum
excitation along a spin chain or the robust transport of an electron
in a quantum dot array, the issue of quantum information transfer is
only partially touched. Indeed, the introduction of an external
field will change the energy of the spin chain system and hence will
necessarily introduce extra dynamical phases to the evolving quantum
system. This makes it clear that transporting quantum phases along
the spin chain requires additional considerations. This quantum
phase issue will be considered in detail in Sec. V.

\section{Adiabatic Transport by a Moving Potential: Computational Results}
In this section, we illustrate our adiabatic quantum population
transfer scheme with detailed computational examples. Let us assume
that the initial state of the spin chain is given by
$|\Phi\rangle=\sum_{m=0}^{N}c_{m}(0)|\bf {m}\rangle$. Two types of
$c_{m}(0)$ will be considered below. In the first case only the
$m=0$th spin is excited, with $c_{m}(0)=\delta_{m0}$. In the second
case, the initial state is a Gaussian wavepacket truncated to three
sites only, with $c_{m}(0)\propto \exp[-(m-1)^{2}/2l_{0}^{2}]$ for
$m=0-2$ and $l_{0}=0.707$ being the width of the Gaussian
wavepacket. In either case a parabolic magnetic field first centered
on the $n=0$th site is applied and then slowly moved to the regime
of larger $n$. This is realized by introducing the time dependence
of $n_{0}$ via $n_{0}=0+St$, where $S$ is the moving speed. Note
that a static parabolic field was previously introduced to induce a
quasi harmonic lower energy spectrum such that good transfer of
Gaussian wavepackets \cite{parabolic} may be realized. By contrast,
our moving potential scenario is more active and effective in
controlling the quantum transport and is in principle applicable to
cases where the shape of the external potential is not parabolic.


The quantum state of the spin chain at a later time can be directly
calculated using the Schr\"{o}dinger equation given above.
In particular, the probability of transferring the quantum
excitation to the last spin of the chain can be examined. If the
performance of the population transfer is satisfactory, one should
find $|c_{N}|^2 \approx 1.0$. Evidently, this condition of high
transfer probability is already useful by itself for, e.g.,
transporting electrons in a quantum dot array in a controlled
fashion.  As shown in Sec. V, the phase of the quantum amplitude
$c_{N}$ may be also taken care of by considering a dual spin chain.

\begin{figure}
  \begin{center}
  \epsfig{file=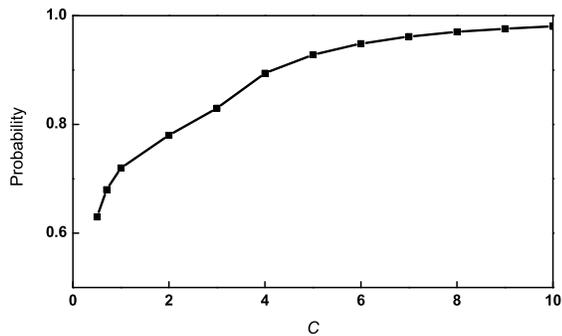,width=8.6cm}
  \end{center}
  \caption{Excitation probability transferred to the last spin in adiabatic quantum transport
  along a chain of 101 spins,
   as a function of the amplitude of the external moving magnetic
   potential characterized by $C$ [in units of $J$, see the text below Eq. (1)].
   The moving speed of the control field is $S=0.005$.}
   \label{cchange}
\end{figure}

We now discuss the feasibility of adiabatic quantum population
transfer by taking advantage of the separatrix associated with the
pendulum analogy. In the ideal case of adiabatic following, an
initial quantum state enclosed by a separatrix will move with the
slowly moving separatrix.  Consider first an initial state localized
exclusively at the $n=0$th site. Then the associated
$k$-distribution covers uniformly from $0$ to $2\pi$. From a
semiclassical perspective afforded by the pendulum analogy, such an
initial state corresponds to an initial ensemble lying on the
$(p-p_{0})=0$ axis of the classical phase space. This initial
ensemble hence necessarily intersects with the separatrix (see Fig.
\ref{separatrix}).  Because the motional period associated with the
separatrix is infinity, those ensemble components that overlap with
the separatrix will always regard the movement of the separatrix as
``too fast to follow". That is, as we slowly move the separatrix
upwards in the classical phase space, some portion of the initial
ensemble may break the adiabaticity and tunnel through the
separatrix structure. Under such a situation adiabatic quantum
population transfer is expected to partially break down. To reduce
the degree of non-adiabaticity, one possible approach is to reduce
the overlap of the initial state with the classical separatrix. This
should be doable by increasing the effective Planck constant
$\sqrt{C}$ (i.e., increasing the strength of the parabolic field)
such that the separatrix regime supports less quantum states. This
is indeed what we find computationally for a chain of 101 spins. In
particular, Fig. \ref{cchange} shows that for a field amplitude
characterized by $C=0.5$, the probability of transferring the
initial excitation to the last spin is only $0.63$. By increasing
$C$ to $8.0$, a transfer probability around $99\%$ is observed.
Figure 3 shows the actual excitation profile $|c_{n}|^2$ vs. $n$ for
a spin chain subject to a parabolic potential moving at a constant
speed of $S=0.005$. At $t=0$, the state is at site $n=0$.
 At $t=10000$, the quantum population is mainly at $n=50$. Note that at that
 moment the excitation profile is slightly delocalized into three sites, but the
 peak probability is still as high as $0.97$. This peak is
 propagated to site $n=100$ at time $t=20000$, with no further
 dispersion detected.  This indicates that our moving
 potential scheme has the capacity to overcome the dispersion issue in
 quantum information transfer.  Though the required field
 strength for large $|n-n_{0}|$ could be demanding experimentally, we point out that
 because the spin excitation is highly localized throughout the
 process, the moving parabolic magnetic field does not need to span
 over many spin sites (in our numerical experiments, we use a parabolic field that
 only spans 20 sites).

\begin{figure}
  \begin{center}
  \epsfig{file=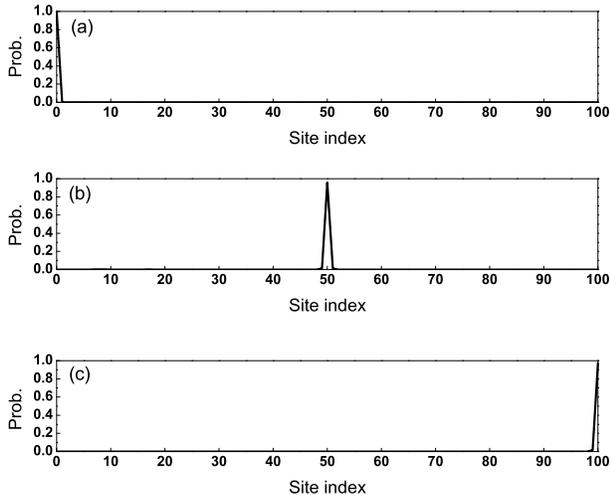,width=9.2cm}
  \end{center}
  \caption{ Adiabatic transfer of spin excitation initially localized exclusively at the $n=0$th site
  along a chain of $101$ spins, for a field amplitude given by $C=8$ and
  its moving speed given by $S=0.005$. Panels (a), (b), and (c) are
  for times $t=0$, $10000$, and $20000$.  Note that the final peak excitation probability remains as high as 0.97.}
\label{single-case}
\end{figure}

\begin{figure}
  \begin{center}
  \epsfig{file=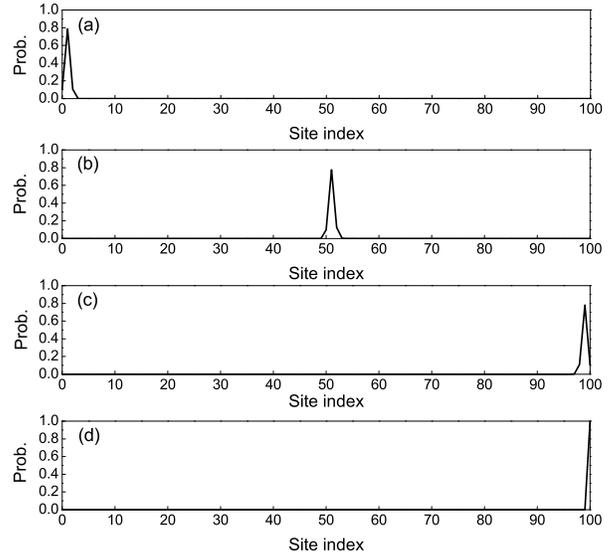,width=9.cm}
  \end{center}
  \caption{Adiabatic quantum transport along a chain of $101$ sites
  with an initial Gaussian excitation profile, at (a) $t=0$, (b) $t=10000$, (c) $t=20000$, and (d)
  $t=21000$. The amplitude of the parabolic field is given by $C=2$, and the moving speed of the control field is
  given by $S=0.005$. Note that the final excitation probability transferred to the last spin is as high as
  0.997.}
  \label{Gaussian-case}
\end{figure}

We have also examined the quantum dynamics for the second type of
initial states, i.e., states with a Gaussian excitation profile at
$t=0$. Because such initial ensembles are localized in both $k$ and
$n$, they can be naturally enclosed by the separatrix shown in Fig.
\ref{separatrix}. As such, if the shape of the initial excitation
profile is appropriately adjusted, the initial state can be made not
to intersect with the separatrix. This being the case, the adiabatic
following should work better, probably requiring a much weaker
parabolic field.
This expectation is also confirmed computationally. In particular,
 Fig. \ref{Gaussian-case} shows the transport of an initial Gaussian excitation
 profile, again for a chain of $101$ spins. At $t=0$, the excitation profile
 spans only the first three sites with a probability peak $0.78$ at the
 site $n=1$. This state is then transported by applying a parabolic field
 with $C=2$ moving at a rate of $S=0.005$. During the quantum transport the
 state disperses among about five sites, with
 a peak probability maintained around $0.77$. At $t=10000$ and $t=20000$, the peak of the spin
 excitation probability profile is transferred
 to $n=50$ and $n=100$. Interestingly, it is found that the
 population transfer probability to the last spin can be further
 enhanced at a slightly later time. As seen from Fig.
 \ref{Gaussian-case}, at time $t=21000$, the peak
probability, located at the last spin, is as high as $0.997$.
Physically, this is due to the reflection process at the end of the
spin chain. In some sense, the interplay of the parabolic field
centered at the end of the spin chain and the reflection process
acts as a lens refocusing the slightly dispersed profile, and the
peak probability builds up on the last spin. Note also that in the
absence of the control field, one in general needs initial
wavepackets of much larger length to be able to reduce the undesired
dispersion \cite{osborne1}.

Results in Fig. \ref{Gaussian-case} demonstrate that using initial
states whose excitation profile covers a few spins can significantly
reduce the required field amplitude (compare values of $C$ in Fig.
\ref{Gaussian-case} and in Fig. \ref{single-case}).
Counter-intuitively, there exists a maximal number $n_{\text{max}}$
of spins that can be used to create such initial states. This can be
appreciated by considering again the separatrix in the classical
phase space of the pendulum analogy. Classical orbits outside the
separatrix are associated with pendulum's rotational motion (rather
than oscillation). Going back to the spin chain or the tight-binding
system, these states correspond to Bloch oscillations in a ``locally
linear" field.  As the separatrix is slowly moving, these states can
continue their Bloch oscillations in a slowly-varying local field
and hence will not follow the motion of the separatrix in the
momentum space. With this understanding, one may estimate
$n_{\text{max}}$ from the width of the separatrix in the momentum
space. Specifically, for a fixed value of the parameter $C$,
$n_{\text{max}} \sim 4\sqrt{\frac{J}{C}}+1$.  For the numerical
example in Fig. \ref{Gaussian-case}, one obtains
$n_{\text{max}}\approx 4$. This estimate is quite consistent with
the finding that during the population transfer, the moving
wavepacket does not cover more than five sites. This result also
implies that for weaker magnetic fields (smaller $C$), one can use
more spins to form the wavepacket for analogous adiabatic population
transfer.

Because our quantum transport scheme is based upon the adiabatic
following of the spin excitation profile with a moving external
potential, it can stop and relaunch the excitation transfer at any
time with great ease, by simply stopping and restarting the movement
of the external parabolic potential. This is simpler than a recent
approach \cite{gong07} using pulsed magnetic fields, and is also
confirmed in our computational studies (not shown).

\begin{figure}
  \begin{center}
  \epsfig{file=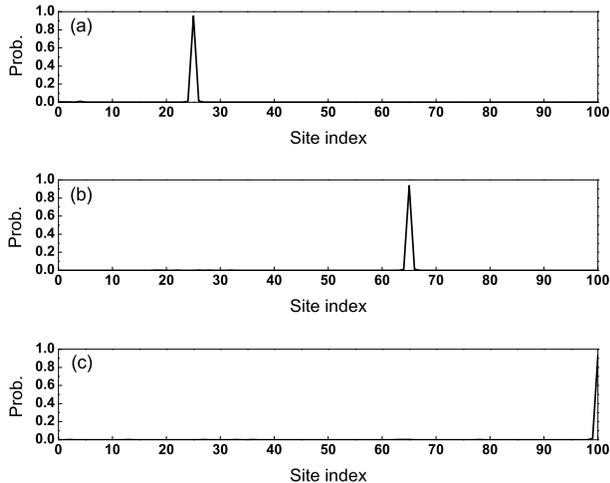,width=9.2cm}
  \end{center}
  \caption{Adiabatic transfer of spin excitation for an initial state exclusively
  localized at the $n=0$th site along a chain of
  $101$ spins. The amplitude of the moving parabolic potential is given by $C=8$, and the moving speed
  is given by $S=0.025$. Panels (a), (b) and (c) show the
  excitation profile at times $t=1000$, $2600$, and $4000$.}
  \label{speed}
\end{figure}

\begin{figure}
  \begin{center}
  \epsfig{file=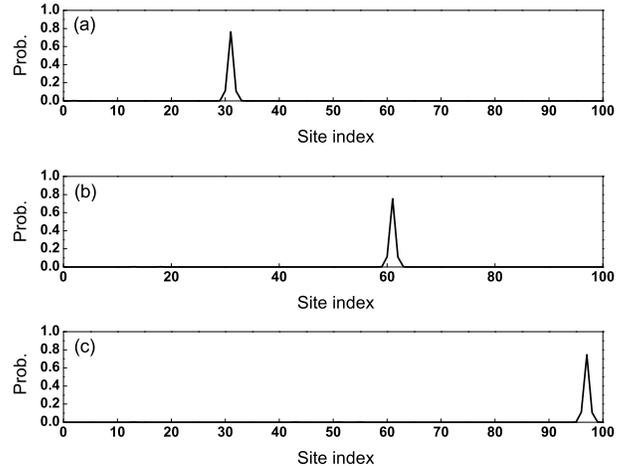,width=9.2cm}
  \end{center}
  \caption{Adiabatic transport of an initial Gaussian profile of spin excitation
   (same as in Fig. \ref{Gaussian-case})
   along a chain of $101$ spins.
The amplitude of the moving parabolic potential is given by $C=2$,
and the moving speed is given by $S=0.1$. Panels (a), (b), and (c)
are for $t=300$, $600$ and $1000$.} \label{speed-Gaussian}
\end{figure}

So can we further increase the moving speed of the control potential
while still maintaining the adiabatic following and hence the
adiabatic quantum population transfer?  Our findings in this regard
can be summarized as follows: (i) for large $C$ adiabatic quantum
transport may survive for a moving speed around $10\%$ of the
coupling constant $J$. The smaller the field strength $C$ is,  the
lower the threshold moving speed will be; (ii) when the moving speed
exceeds the threshold, the probability of successful population
transfer gradually decreases,  but can still be considerably large
for a relatively short spin chain. For example,  Fig. \ref{speed}
shows the result for an initial state exclusively localized at the
$n=0$th site. The moving speed of the parabolic potential is
$S=0.025$. The peak value of the probability profile is $0.96$ at
time $t=1000$. It reduces to about $0.95$ at $t=2600$ and $0.94$ at
$t=4000$.  Hence, in this case, only $2\%$ reduction in the peak
probability occurs when the moving speed $S$ increases by a factor
of five. However, increasing the moving speed beyond this limit
drastically reduces the probability of population transfer to the
last spin. For a moving speed of $S=0.30$, the probability maxima
equals only $0.87$ at $t=2200$, and $0.84$ at $t=3400$. Analogous
calculations are also carried out for the transport of an initial
Gaussian excitation profile. As shown in Fig. \ref{speed-Gaussian},
for a moving speed of 0.1, the peak value of probability remains
around $0.76$ during the transport process. As such, at the end (not
shown) the adiabatic population transfer is also very successful for
this high moving speed. But if the moving speed is further increased
by several times, dispersion in the spin excitation profile will
be considerable. 

In short, our numerical experiments suggest that, to achieve
adiabatic transport of spin excitation along a quite long spin chain
using a moving parabolic potential, the associated moving speed can
be as large as being one tenth of the natural propagation rate ($J$)
of the system (without disorder).

\begin{figure}
  \begin{center}
  \epsfig{file=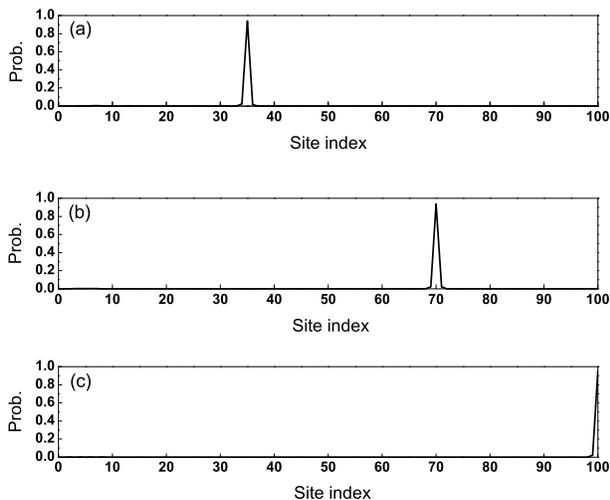,width=9.2cm}
  \end{center}
  \caption{Adiabatic transfer of spin excitation for an initial state
  exclusively
  localized at the $n=0$th site along a chain of $101$ spins, in the presence of static disorder with
  the noise amplitude given by $\Delta =0.5$. Other parameters are the same as in Fig.
  \ref{single-case}. The excitation profile at time $t=7000$, $14000$, and $20000$
   are shown in panels (a), (b), and (c). Results here are very similar to those shown in Fig. \ref{single-case}.}
  \label{single-case-static}
\end{figure}

\section {Robustness of Adiabatic Transport}
So far the majority of quantum state transfer schemes consider only
idealized spin chains with no disorder in the spin-spin coupling
strength. This suggests a gap between theoretical exploration and
realistic situations in experiments. In particular, the effects of
static and dynamic imperfections in spin chains are studied in very
few cases \cite {lyakhov1,burgarth1,burgarthnj,disorder}. Here we
computationally study the influence of static and dynamic disorder
on our adiabatic population transfer scheme, by considering the
model Hamiltonian in Eq. (\ref{KReq}) with fluctuating spin-spin
coupling constants.  We hope to numerically confirm the robustness
of our scheme as implied by its adiabatic nature.

The model Hamiltonian with disorder in the spin-spin coupling
strength is given by
\begin{eqnarray}
  H_{sd} &=&\sum_{n=0}^{N-1}-\frac{(J+\delta_{n})}{2}{\bf \sigma}_{n}\cdot {\bf
  \sigma}_{n+1} \nonumber \\
 &&
   +\sum_{n=0}^{N}\frac{C}{2}(n-n_{0})^2\sigma_{n}^{z},
\end{eqnarray}
where $\delta_{n}$ are time-independent random numbers uniformly
distributed in the interval  $[-\Delta,\Delta]$, representing random
fluctuations in $J$ with the amplitude $\Delta$.  We call cases with
such  time-independent disorder as static disorder models. Note that
specific results presented below refer to single disorder
realizations. These results are very typical such that there is no
need to average over many disorder realizations.

Figure \ref{single-case-static} display one sampling calculation
that is in parallel with the results in Fig. \ref{single-case} but
takes into account static disorder with the noise amplitude
$\Delta=0.5$.  As demonstrated in Fig. \ref{single-case-static},  in
the presence of such a high noise level,  the quantum population of
spin excitation is still successfully transferred to the last spin
of the chain (peak probability around 99\%), with the excitation
profile almost unaltered as compared with the noiseless case studied
in Fig. \ref{single-case}. For an even higher fluctuation level,
e.g., $\Delta=0.7$, the spin excitation profile is seen to gradually
disperse as it is transported along the chain.

\begin{figure}
  \begin{center}
  \epsfig{file=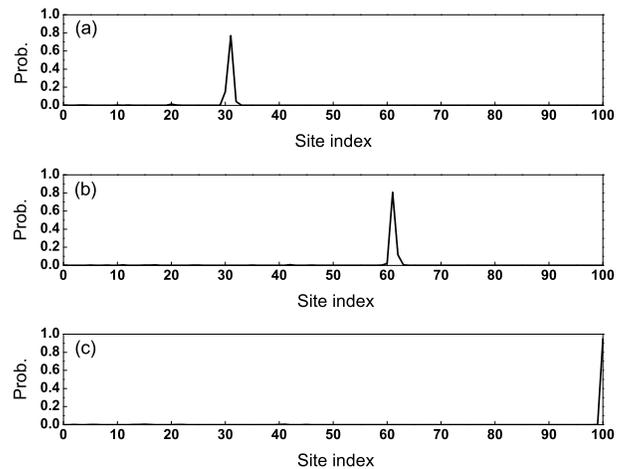,width=9.2cm}
  \end{center}
  \caption{Adiabatic transfer of an initial Gaussian profile of spin excitation along
  a chain of $101$ spins, in the presence of static disorder
  characterized by $\Delta=0.5$. Other parameters are the same as in Fig. \ref{Gaussian-case}.
  Panels (a), (b) and (c) are for
$t=6000$, $12000$ and $21500$. Results here are very similar to
those shown in Fig. \ref{Gaussian-case}.}
\label{Gaussian-case-static}
\end{figure}

Figure \ref{Gaussian-case-static} displays results for an initial
Gaussian excitation profile also considered in Fig.
\ref{Gaussian-case}, but in the presence of static disorder
characterized by $\Delta=0.5$. We find that the severe disorder can
slightly change the shape of the spin excitation profile (though not
so evident in Fig. 8) and hence the peak value of the probability
profile slightly fluctuates during the controlled transport. Note
however, the area enclosed by the main probability profile is found
to be around $0.99$ at all times. At time $t=21500$, the peak
probability of the spin excitation profile, still as large as
$0.99$, has been transferred to the $100$th site as in the noiseless
case of Fig. \ref{Gaussian-case}. In another sampling case for a
$\Delta=0.7$, the peak excitation probability that is transferred to
the last spin decreases to $0.94$. All these results clearly
demonstrate the robustness of our adiabatic transport scheme to
high-level static disorder.

We have also examined the robustness of our adiabatic transport
scheme to dynamic disorder. To model time-dependent fluctuations in
the spin-spin coupling strength, we now let each $\delta_{n}$ be
given by the sum of ten oscillating functions, i.e.,
$\delta_{n}= \sum_{i=1}^{10} A\cos(\omega_{i} t+\phi_{i})$, where
$\omega_{i}$ are random frequencies distributed in
$[0,\omega_{\text{max}}]$, and $\phi_{i}$ are random phases
uniformly distributed in $[0,2\pi]$.

\begin{figure}
  \begin{center}
  \epsfig{file=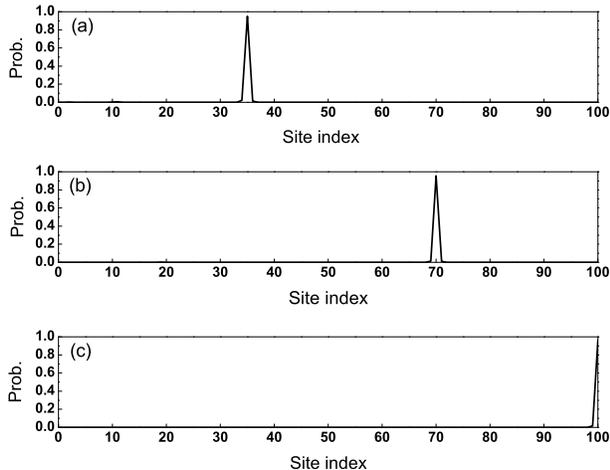,width=9.2cm}
  \end{center}
  \caption{Adiabatic transport of spin excitation for an
  initial state localized exclusively at $n=0$th site along a chain of $101$ spins,
  in the presence of dynamic disorder
  characterized by
$A=0.025$ and $\omega_{\text{max}}=0.1$. Other parameters are the
same as in Fig. \ref{single-case-static}. Panels (a), (b), and (c)
are for $t=7000$, $14000$ and $20000$.} \label{single-case-dynamic}
\end{figure}

\begin{figure}
  \begin{center}
  \epsfig{file=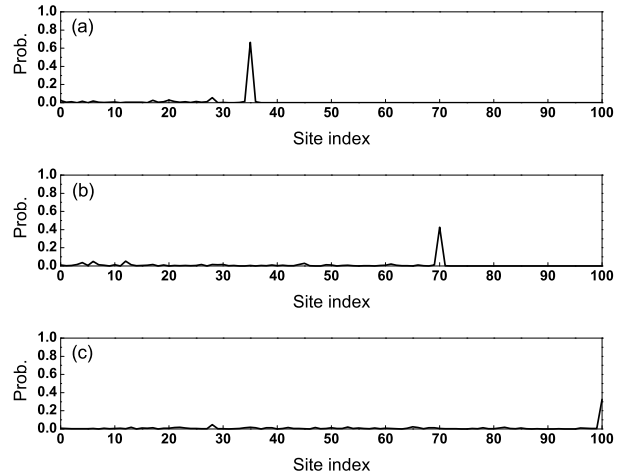,width=9.2cm}
  \end{center}
  \caption{Adiabatic transfer of spin excitation for an intial state exclusively localized
   at the $n=0$th site along a chain of $101$ spins. $\omega_{\text{max}}=1.0$, and other
   parameters are the same as in Fig.
   \ref{single-case-dynamic}.}
   \label{single-case-dynamic2}
\end{figure}

Interestingly, our numerical experiments indicate that effects of
dynamical disorder modeled above depend strongly on
$\omega_{\text{max}}$, i.e., the cut-off frequency of the dynamic
fluctuations. Introducing disorder more frequently, i.e.,
introducing a larger $\omega_{\text{max}}$, can lead to much
decreased peak probability transferred to the last spin.  In
particular, we find that for $\omega_{\text{max}}\leq 0.1 $, the
effects of the dynamic disorder are essentially analogous to what is
found for static disorder.  For larger $\omega_{\text{max}}$, the
deterioration of the adiabatic population transfer becomes
considerable for the same noise amplitude $A$. Figure
\ref{single-case-dynamic} displays the results for $A=0.025$ and
$\omega_{\text{max}}=0.1$. Note that for $A=0.025$, the amplitude of
the noise is very large because the total fluctuation is a sum of
ten functions oscillating at the same amplitude $A$. It is seen from
Fig. \ref{single-case-dynamic} that for such a case of dynamic
disorder, the spin excitation travels almost unaffected along the
chain, thus confirming again the robustness of our adiabatic scheme.
However, upon an increase in $\omega_{\text{max}}$, e.g.,
$\omega_{\text{max}}=1.0$ (so noise frequency becomes comparable to
the characteristic coupling strength $J$), the dispersion of the
spin excitation profile becomes evident in Fig.
\ref{single-case-dynamic2}. The situation can be certainly much
improved if the noise amplitude $A$ is decreased. Because similar
results are also found for Gaussian excitation profile as initial
states, we conclude that the noise spectrum of dynamic disorder can
play an important role in affecting the robustness of our adiabatic
transport scheme, especially when the noise amplitude is very large.

Although our adiabatic scheme is seen to be robust, it can be
expected that the very existence of disorder should limit the
threshold speed of the moving potential. Put alternatively, for a
larger moving speed (which satisfies the adiabatic condition less),
the robustness of our control scheme to disorder is expected to
decrease. This trend is indeed found in our numerical experiments.
To characterize precisely how an increasing moving speed of the
control potential affects the robustness is certainly beyond the
scope of this work.

\section{Adiabatic Transport in a Dual Spin Chain}

As demonstrated in previous sections, our adiabatic scheme based
upon a moving parabolic potential offers a simple and robust
approach to transferring quantum population along a spin chain. This
scheme requires a strong parabolic field if it is globally
parabolic, but even this requirement can be greatly weakened if the
initial spin excitation profile spans a few sites. Further, one does
not really need a globally parabolic field to realize this adiabatic
scheme: it suffices for the parabolic field profile to be wider than
the spin excitation profile.  With these considerations we may argue
that the well-known dispersion issue in quantum information transfer
along spin chains is essentially solved by our adiabatic scheme.
Nevertheless, as also mentioned earlier, one important issue still
remains open. That is, for the sake of quantum information transfer,
how to take care of the quantum phase of a quantum state to be
transported? Indeed, a moving external potential induces extra
dynamical phase to the spin chain, and such a dynamical phase
depends on the details of the control potential. These facts
motivate us to seek an encoding approach that can protect a quantum
state from the additional dynamical phases induced by the moving
potential.

Fortunately, the idea of using a dual spin chain, first proposed to
overcome the disorder and dispersion issues in quantum information
transfer in spin chains \cite{burgarth2,burgarthnj}, offers a
promising solution. Specifically, we propose to combine our
adiabatic transport scheme with the dual spin chain scheme. Then,
because each individual sub-chain acquires identical dynamical
phases from the same external moving potential, the relative quantum
phase between the two sub-chains is certain, and hence quantum
information encoded in the dual spin chain can be transported
without suffering from the extra uncertain dynamical phases.

Consider then a quantum channel consisting of two identical parallel
spin chains subject to the same external parabolic potential,
\begin{eqnarray}
  H^{i}_{s} &=&-\frac{J}{2}\sum_{n=0}^{N-1}{\bf \sigma}^{(i)}_{n}\cdot {\bf
  \sigma}^{(i)}_{n+1} \nonumber \\
&& +\sum_{n=0}^{N}C\frac{(n-n_{0})^2}{2}\sigma_{n}^{z(i),}
\end{eqnarray}
where $i=1,2$ indices label the two sub-chains. Suppose the quantum
state to be transferred is given by $|\Phi\rangle = \alpha|0\rangle
+ \beta|1\rangle$. Such a state can be encoded into the quantum
channel prepared in the following entangled state,
\begin{eqnarray}
 |\mathbf{\Psi}(0)\rangle &=&\alpha|\mathbf{g}\rangle^{(1)}\otimes|\mathbf{0}\rangle^{(2)}+\beta|\mathbf{0}\rangle^{(1)}
 \otimes|\mathbf{g}\rangle^{(2)}
 \label{initialstate}
\end{eqnarray} as a superposition of two components:
the $n=0$th spin in the second (first) sub-chain being flipped and
the first (second) sub-chain in its ground state denoted by
$|\mathbf{g}\rangle$.  Note that for each sub-chain at most one spin
is flipped and the associated dynamics will be restricted to the
ground state or the subspace of one flipped spin.

This encoding can be extended to cases of entangled Gaussian
wavepackets in a straightforward manner. However, for convenience
here we discuss only cases arising from the initial state given by
Eq. (\ref{initialstate}).  After the independent evolution of the
two sub-chains for a total duration of $\tau$ under the action of
the moving parabolic potential, the quantum state of the dual spin
chain is given by
\begin{eqnarray}
  |\Psi(\tau)\rangle &=&\sum_{n=0}^{N}c_{n}(\tau)|\Phi_{n}\rangle,
\end{eqnarray}
where $|\mathbf{\Phi}_{n}\rangle \equiv
\alpha|\mathbf{g}\rangle^{(1)}\otimes|\mathbf{n}\rangle^{(2)}+\beta|\mathbf{n}\rangle^{(1)}
\otimes|\mathbf{g}\rangle^{(2)}$.  Evidently, though each $c_{n}$
contains the extra quantum phases induced by the external moving
potential, this factor is identical for the two state components of
$|\mathbf{\Phi}_{n}\rangle$.  As already demonstrated in our
numerical experiments using a single spin chain, the profile of
$|c_{n}|^{2}$ should also be highly localized, and the time of
arrival of the peak value of $|c_{n}|^{2}$ at the last spin can also
be directly calculated from the moving speed of the parabolic
potential.

Analogous to the original dual spin chain scheme, at the end of the
adiabatic quantum transport the final state $|\Psi(\tau)\rangle$ can
be decoded by applying a CNOT operation  to the last two $N$th spins
of the dual chain. Upon this operation the final state is
transformed to
\begin{eqnarray}
\sum_{n=0}^{N-1}c_{n}(\tau)|\Phi_{n}\rangle  + c_{N}(\tau)\left[
\alpha |\mathbf{g}\rangle^{(1)}+\beta
|\mathbf{N}\rangle^{(1)}\right]\otimes|\mathbf{N}\rangle^{(2)}.
\end{eqnarray}
As such, by measuring the last spin of the second sub-chain, one
gains important information about the transport. In particular, if
the measurement outcome is spin up, then the initial state
$|\Phi\rangle = \alpha|0\rangle + \beta|1\rangle$ has been
successfully transferred to the last spin of the first sub-chain,
with probability $|c_{N}(\tau)|^{2}$; if the outcome is spin down,
then the quantum state transfer is unsuccessful and one needs to
wait for more time to perform additional measurements.

Significantly, because our adiabatic population transfer scheme can
ensure a very high probability of excitation transfer to the last
spin, the probability of spin-up measurements can be guaranteed to
be very high (arbitrarily high if there were no restriction on the
field strength). This hence overcomes, at least theoretically, one
main disadvantage of previous dual spin chain schemes where too many
measurements may be required for high fidelity quantum state
transfer. Further, at the end of the adiabatic transport, the spin
excitation is automatically localized at very few end spins. So it
also becomes unnecessary to perform fast measurements at a
particular time. Instead, one can choose measurement times at will
so long as the moving parabolic potential has reached the last site
of the spin chain. This makes it clear that our adiabatic scheme,
when combined with quantum phase encoding schemes, can find
important applications in quantum information transfer (in addition
to quantum population transfer).

\section{Conclusions}
In this work we have presented a simple and robust scheme to realize
adiabatic population transfer in spin chains. The additional
resource needed is a slowly moving external parabolic magnetic
field.  The basic mechanism is the adiabatic following of a quantum
state with the movement of a separatrix structure in the classical
phase space of a pendulum analogy.  In particular, we have shown
that our scheme can be used to transfer spin excitation from one end
of a spin chain to the other end, with the initial excitation
profile being a localized truncated Gaussian wavepacket or
exclusively localized at a single spin site. It is found that much
weaker external field is needed for adiabatic population transfer if
the initial excitation profile covers a few spin sites. Effects of
static and dynamical fluctuations in the spin-spin coupling strength
are also computationally studied, confirming the robustness of our
adiabatic population transfer scheme. Realizing the robust
population transfer with small dispersion, we have also proposed to
apply our approach to a dual spin chain such that robust quantum
information transfer can be realized with important advantages. We
hope that our theoretical scheme can motivate experiments using
various implementations of a spin chain Hamiltonian, such as cold
atoms in an optical lattice and electron tunneling in an array of
quantum dots.

The central idea of this work, namely, using a slowly moving
external potential to adiabatically transfer spin excitation, might
be useful for other applications as well. For example, one may
consider distributing entanglement along a long spin chain in a
controlled fashion, by use of a control potential that has two
components slowly moving in opposite directions. Another interesting
application is related to studies of quantum signal amplification
with spin chain models. Recently, an interesting connection between
quantum state transfer and quantum state amplification is revealed
\cite{kay}. In this regard, our adiabatic scheme might also help
design a new and useful approach to controlled quantum amplification
using a slowly moving external potential.


\acknowledgments

 J.G. is supported by the start-up funding, (WBS grant No.
R-144-050-193-101 and No. R-144-050-193-133), National University of
Singapore, and the NUS ``YIA" funding (WBS grant No.
R-144-000-195-123) from the office of Deputy President (Research \&
Technology), National University of Singapore.

        \end{document}